\documentclass[manuscript]{aastex}

\shorttitle{LFQPO in GRS~1915+105 $\rho$ State}
\shortauthors{Yan et al.}

\usepackage{times}
\input{epsf}
\usepackage{epsfig}
\usepackage{longtable}
\usepackage{natbib}
\begin{document}

\title{Phase-Resolved Timing Analysis of GRS~1915+105 in Its $\rho$ State}

\author{Shu-Ping Yan\altaffilmark{1,2}, Na Wang\altaffilmark{1}, Guo-Qiang Ding\altaffilmark{1}, and Jin-Lu Qu\altaffilmark{3}}

\altaffiltext{1}{Xinjiang Astronomical Observatory, Chinese Academy of Sciences, 150, Science 1-Street, Urumqi, Xinjiang 830011, China; yanshup@xao.ac.cn, na.wang@xao.ac.cn}
\altaffiltext{2}{University of Chinese Academy of Sciences, 19A Yuquan road, Beijing 100049, China}
\altaffiltext{3}{Key Laboratory for Particle Astrophysics, Institute of High Energy Physics, Chinese Academy of Sciences, 19B Yuquan Road, Beijing 100049, China}

\begin{abstract}

We made a phase-resolved timing analysis of GRS~1915+105 in its $\rho$ state and obtained detailed $\rho$ cycle evolutions of the frequency, the amplitude and the coherence of low-frequency quasi-periodic oscillation (LFQPO). We combined our timing results with the spectral study by Neilsen et al. to perform an elaborate comparison analysis. Our analyses show that the LFQPO frequency does not scale with the inner disk radius, but it is related to the spectral index, indicating a possible correlation between the LFQPOs and the corona. The LFQPO amplitude spectrum and other results are naturally explained by tying the LFQPO to the corona. The similarities of the spectra of variability parameters between the LFQPO from $\rho$ state and those from more steady states indicate that the LFQPOs of GRS~1915+105 in very different states seem to share the same origin.

\end{abstract}


\section{Introduction}

GRS~1915+105 is a binary system that was discovered by WATCH on board {\emph{GRANAT}} in 1992 \citep{Castro92}. It is located in our Galaxy at an estimated distance of about $11$ kpc \citep[e.g.,][]{Fender99, Zdziarski05}, containing a spinning black hole \citep {Zhang97, McClintock06} with mass $14\pm4$ $M_{\odot}$, and a K-M III giant star with mass $0.8\pm0.5$ $M_{\odot}$ as the donor \citep{Harlaftis04, Greiner01b}. The orbital separation of the binary components is about $108\pm4$ $R_{\odot}$ and the orbital period is $33.5\pm1.5$ days \citep{Greiner01a}. As the first microquasar found, GRS~1915+105 produces superluminal radio jets \citep{Mirabel94, Fender99}. The count rate and color characteristics are extremely complex and the light curves of the source are usually classified into 12 variability classes which are regarded as transitions between three basic states, states A, B and C \citep{Belloni00}. Lots of low-frequency ($\sim$0.5--10 Hz) quasi-periodic oscillations (LFQPOs) are found in GRS~1915+105, providing us ideal subjects for the study of LFQPO.

Fruitful results have been obtained from the LFQPO of GRS~1915+105. It is revealed that the LFQPO frequency is positively correlated with the fluxes of the individual components of the spectra, i.e., the thermal and power-law components or the source intensity \citep[e.g.,][]{Chen97, Markwardt99, Muno99, Muno01, Trudolyubov99, Reig00, Tomsick01}. The LFQPO amplitude is inversely correlated with the source flux or LFQPO frequency \citep[e.g.,][]{Muno99, Reig00, Trudolyubov99}. As the LFQPO frequency increases, the temperature of the inner accretion disk increases and the radius of the inner accretion disk decreases \citep{Muno99, Rodriguez02a}. The LFQPO amplitude increases with photon energy and in some cases it turns over in the higher energy bands \citep[e.g.,][]{Tomsick01, Rodriguez02b, Rodriguez04, Zdziarski05}. As the centroid frequency of the LFQPO increases, the relation between the LFQPO frequency and photon energy evolves from a negative correlation to a positive one \citep{Qu10}. In addition, three additional combined patterns of the negative correlation and the positive one were discovered in a systematic study of the LFQPO frequency$-$photon energy relationship \citep{Yan12}.

Although the results mentioned above enable a good understanding of LFQPO phenomenology, we are puzzled by the ambiguity that the LFQPO is correlated with both the corona/jet and accretion disk. The LFQPO origin is still a mystery. \citet[hereafter NRL11]{Neilsen11} investigated the physical changes and the LFQPO evolution of the $\rho$ variability in GRS~1915+105 through a phase-resolved spectral and timing analysis. In order to reveal more clues about the origin of the LFQPO and more details about evolution of the $\rho$ cycle, in this work we make a detailed comparison between the results of spectral and timing analyses and investigate the linkage between the $\rho$-associated LFQPOs and the LFQPOs from GRS~1915+105 during more steady conditions. In addition, we attempt to use the ratio of the power density spectrum (PDS) continuum amplitude to the LFQPO amplitude to track the origin of aperiodic X-ray variability from accretion flows.

The observation and data reduction methods are described in Section 2, the results are presented in Section 3, and the discussion and conclusion are given in Section 4.

\section{Observation and Data Reduction}

With a method simialr to that of NRL11, we produce the average phase-folded light curve for 60405-01-02-00, the same {\it RXTE} observation of GRS~1915+105 that was analyzed by NRL11.

We perform our analyses with the HEASOFT version 6.10 package. The light curves with a time resolution of 1 s are extracted from the binned mode data (B\_8ms\_16A\_0\_35\_H\_4P) in 2.0--14.8 keV and the event mode data (E\_16us\_16B\_36\_1s) in 14.8--60 keV. The data are accumulated from four PCUs in combinations of \{0, 2, 3, 4\} and \{1, 2, 3, 4\}. Good time intervals are defined as follows: a satellite elevation over the Earth limb $>10^{\circ}$ and an offset pointing $<0.02^{\circ}$. The rows with non-common time are deleted from the two light curve FITS files. Then, we add the two dead-time-corrected and background-subtracted light curves and obtain a light curve that will be further barycentered.

We obtain a series of cycles from the barycentered light curve and then cross-correlate them with a representative cycle. Maxima in the normalized cross-correlation values correspond to the peaks in the light curve. We smooth the cross-correlation with a Gaussian of FWHM 3 s and measure the time of maxima via parabolic interpolation. The resulting maxima are phase $\phi=0$ times which are used to create the average phase-folded light curve. In order to mitigate the effect of the initial template choice, we use the folded light curve as a new template and cross-correlate it with the entire set of cycles and obtain the final set of $\phi=0$ times.

For the timing analysis, the light curve in a certain phase interval is extracted with a time resolution of 8 ms. The PDS is produced with the dead-time-corrected Poisson noise level subtracted from the PDS \citep{Morgan97} and with the normalization of \citet{Miyamoto92}, which gives the periodogram in units of (rms/mean)$^2$/Hz. The PDS for each 0.02 phase and 0.4--0.74 phase are computed on 2 s and 4 s sampling durations, respectively. Following \citet{Belloni02}, the PDS is fitted with a model that includes several Lorentzians to represent the continuum, the QPOs, and other broad features, respectively. The fitting is limited over the 0.5--20 Hz frequency band. 

In order to study the spectra of the variability parameters (e.g., continuum amplitude, LFQPO amplitude, LFQPO frequency), we produce a PDS for several energy bands (see Figure \ref{fig:lanRms2e2Fre}). The continuum amplitudes in several frequency bands are obtained by subtracting the LFQPO contributions (the LFQPO parameters are obtained from the fitting mentioned above) from the full integrals in corresponding frequency bands \citep[see, e.g.,][]{Vaughan03}. Due to the high energy dependence of the background, when studying the energy-dependent PDS, we correct the continuum amplitude and LFQPO amplitude for background following $A_{\rm net}=A_{\rm raw}\times\frac{S+B}{S}$, with $A$ being the amplitude, $S$ the source net rate, and $B$ the background rate \citep{Berger94, Rodriguez11}. The errors are derived by varying the parameters until $\Delta\chi^2=1$, at 1 $\sigma$ level.

\section{Results}

\subsection{Timing Analysis for 0.02 Phase Intervals}

\begin{figure}
\centerline{\includegraphics[height=10cm,angle=-90]{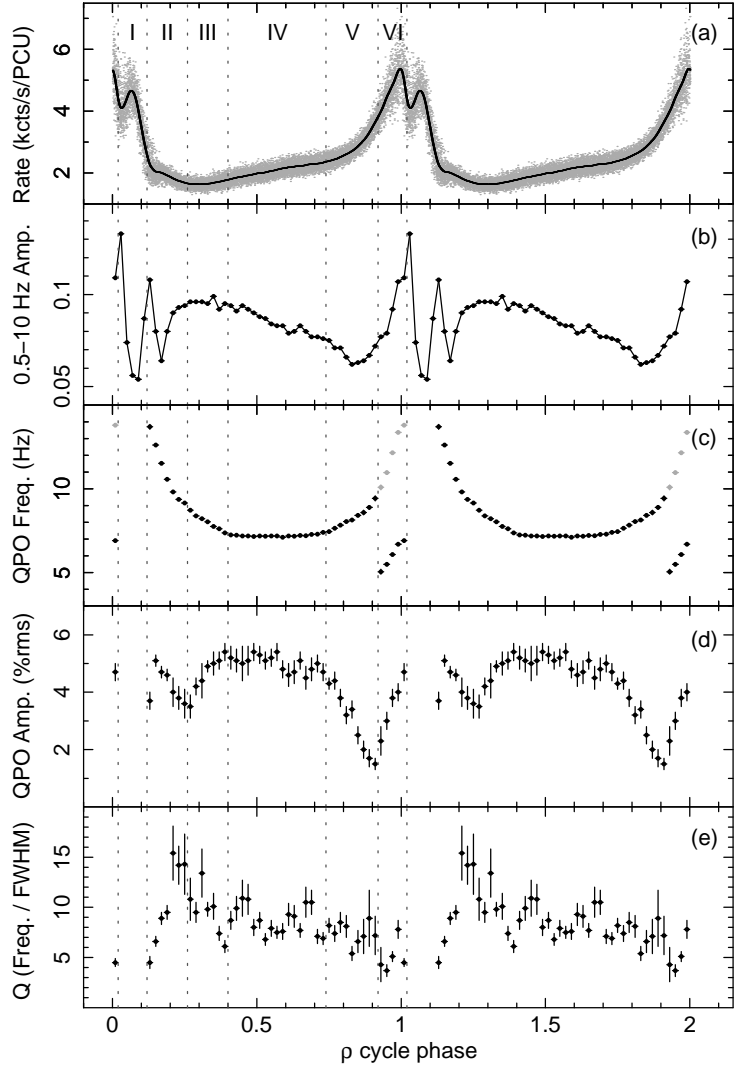}}
\caption{\label{fig:lan.LFQPO}(a) The phase-folded PCA $\rho$ class light curve in 2.0--60 keV band. (b) the 0.5--10 Hz rms amplitude, (c) LFQPO frequency, (d) LFQPO amplitude, and (e) LFQPO coherence, as functions of $\rho$ cycle phase. 
In interval I, there is no obvious LFQPO. As phase increases, the LFQPO frequency decreases (II and III), flattens (IV) and then increases again (V). We double the LFQPO frequencies in interval VI and find that they (the gray points) and the other LFQPO points together show a much more symmetric time evolution over the cycle. In intervals II and VI, the LFQPO amplitude is positively correlated with the LFQPO frequency. While in intervals III, IV and V, the LFQPO amplitude is negatively correlated with the LFQPO frequency. As phase increases, the coherence of LFQPO reveals a rapid increase (II), and then decreases quickly (III) followed by a continuous gentle decrease (IV, V and VI). The horizontal bars denote phase ranges. The vertical bars are error bars.}
\end{figure}

The PDS for each 0.02 phase interval is computed in the 2.0--37.8 keV band and the results are shown in Figure \ref{fig:lan.LFQPO}. The phase-folded PCA $\rho$ class light curve (Figure \ref{fig:lan.LFQPO}(a)) and the shape of the PDS power integrated over 0.5--10 Hz (hereafter referred to as the 0.5--10 Hz amplitude, and similarly for other amplitudes) (Figure \ref{fig:lan.LFQPO}(b)) are similar to those of NRL11, except that our count rate is higher and the 0.5--10 Hz amplitude (2 s sample duration) is a bit lower than NRL11's 2--10 Hz amplitude (1 s sample duration). Based on the behavior of the LFQPO amplitude (Figure \ref{fig:lan.LFQPO}(d)), the cycle phase is divided into six intervals (I: 0.02--0.12, II: 0.12--0.26, III: 0.26--0.4, IV: 0.4--0.74, V: 0.74--0.92, and VI: 0.92--0.02). In interval I, there is no obvious LFQPO. In intervals II and VI, the LFQPO amplitude is positively correlated with the LFQPO frequency. Meanhile for intervals III, IV and V, the LFQPO amplitude is negatively correlated with the LFQPO frequency. In view of the overall situation, the LFQPO amplitude decreases very slightly in interval IV but dips in phase intervals 0.12--0.4 (II and III) and 0.74--0.02 (V and VI). The LFQPO amplitude in the phase interval 0.12--0.14 is relatively small. As phase increases, the LFQPO frequency decreases (II and III), flattens (IV) and then increases again (V). It is very interesting that the LFQPO frequency remains constant in interval IV, and drops at $\phi=0.92$. We double the LFQPO frequencies in interval VI and find that they (the gray points) and the other LFQPO points together show a much more symmetric time evolution over the $\rho$ cycle (Figure \ref{fig:lan.LFQPO}(c)). As phase increases, the coherence of LFQPO reveals a rapid increase (II), and then a quick decrease (III) followed by a continuous gentle decrease (IV, V, and VI) (Figure \ref{fig:lan.LFQPO}(e)).

The LFQPO absolute amplitude is estimated by multiplying the LFQPO amplitude and source count rate \citep[see, e.g.,][]{Mendez97, Gilfanov03, Zdziarski05}. The LFQPO absolute amplitude has a behavior similar to that of the LFQPO amplitude. It increases very slightly in interval IV but dips in phase intervals 0.12--0.4 (II and III) and 0.74--0.02 (V and VI) (Figure \ref{fig:lan.absrms}). In interval VI, it rises up rapidly.

\subsection{Spectra of Variability Parameters}

Spectra of variability parameters provide us a way to find the correlations between variability and the regions where photons with different energy are produced. They are also good indicators of where to find the connections between the LFQPOs in the $\rho$ state and other states. We produce the spectra of the 0.5--64 Hz, 3--64 Hz, and 10--64 Hz continuum amplitudes, the LFQPO amplitude and the LFQPO frequency for interval IV during which the LFQPO frequency and the power-law index are both relatively steady (Figure \ref{fig:lan.LFQPO}; Figure 7 in NRL11). 

The LFQPO frequency almost increases linearly with photon energy when the energy $\lesssim15$ keV, and then it levels off (Figure \ref{fig:lanRms2e2Fre}(a)). The points are fitted with the least squares \citep[see, e.g.,][]{Greene02}. The slope of the best-fit line (red) is $0.0041\pm0.0010$ Hz keV$^{-1}$ and the adjusted R-square is 0.65. 

Figure \ref{fig:lanRms2e2Fre}(b) shows the 0.5--64 Hz, 3--64 Hz, 10--64 Hz continuum amplitude spectra and the LFQPO amplitude spectrum. As the photon energy increases, all these amplitudes increase and then flatten out. The ratios of the continuum amplitudes to the LFQPO amplitude are shown in Figure \ref{fig:lanRms2e2Fre}(c). As photon energy increases, all of the ratios decrease rapidly and then smoothly level off. The points can be well fitted with the function $R(E)=A-D\,B\ln\{ \exp [(E_{\rm tr}-E)/D]+1\}$ \citep[function (1) in][]{Shaposhnikov07}. We fix the $D$ at 1 keV. The best-fit values for the $0.5-64$ Hz continuum /LFQPO amplitude ratio are $A=1.51\pm0.08$, $B=-0.33\pm0.08$ keV$^{-1}$, and $E_{\rm tr}=7.53\pm0.95$ keV. The fit for the 3--64 Hz continuum/LFQPO amplitude ratio gives: $A=1.26\pm0.09$, $B=-0.26\pm0.11$ keV$^{-1}$, and $E_{\rm tr}=6.99\pm1.43$ keV. The fit for the 10--64 Hz continuum/LFQPO amplitude ratio gives: $A=1.05\pm0.10$, $B=-0.25\pm0.20$ keV$^{-1}$, and $E_{\rm tr}=6.05\pm2.20$ keV. The errors for the best-fit parameters are standard deviations.

\subsection{Spectral Index--LFQPO Frequency Relation}

In order to further investigate the origin of the LFQPO and the evolution of the $\rho$ cycle, the ratio of the spectral index to the LFQPO frequency is presented in Figure \ref{fig:lanGa2Fre}. The spectral index is obtained from Figure 7 in NRL11. In interval IV, both the LFQPO frequency and the power-law index remain almost constant, and the ratio is $\sim0.4$. As the LFQPO frequency increases, the index increases, while the ratio decreases faster. We must note that the ratio is an artificial quantity with arbitrary units and is merely a vehicle to track the correlation between the Compton spectrum and the LFQPO frequency versus phase.

\section{Discussion and Conclusion}

In this section, we combine our timing results described in Section 3 with NRL11's spectral analysis to investigate the origin of the LFQPO from GRS~1915+105 during its $\rho$ state.

In the $\rho$ state, NRL11 showed that the energy spectrum has at least two components: the disk emission and the corona emission. The LFQPO may originate from the accretion disk, the corona, both the disk and the corona, or the interaction of the disk with the corona.

As phase increases, the LFQPO frequency decreases (II and III), flattens (IV) and then increases again (V) (Figure \ref{fig:lan.LFQPO}(c)), while the accretion disk radius increases continuously over the phase range 0.1--0.9, which covers intervals II, III, IV, and V (Figure 7 in NRL11). Assuming that the spectral model in NRL11 is correct, it turns out that the LFQPO frequency does not scale with the inner disk radius, suggesting that the LFQPO cannot be tied to dynamical frequencies in the disk. Although the possibility of the LFQPO originating from disk should not be ruled out based only on this, it is still a puzzle whether the LFQPO originates from the corona.

\begin{figure}
\centerline{\includegraphics[height=12cm,angle=-90]{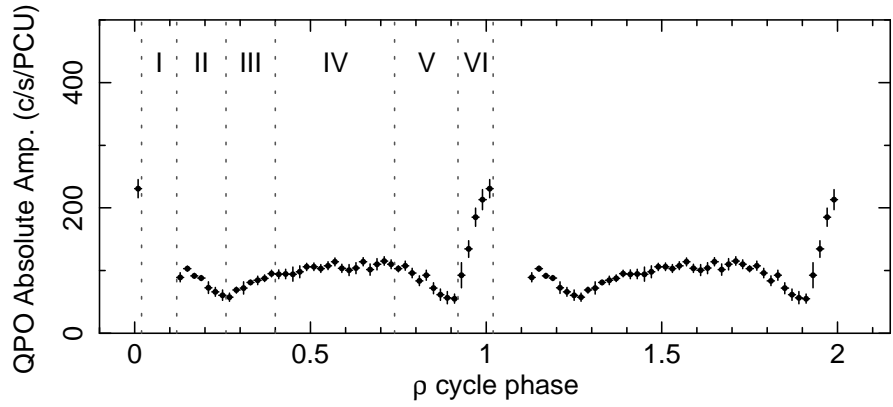}}
\caption{\label{fig:lan.absrms} the LFQPO absolute amplitude as a function of $\rho$ cycle phase. The LFQPO absolute amplitude increases very slightly in interval IV but dips in phase intervals 0.12--0.4 (II and III) and 0.74--0.02 (V and VI). The horizontal bars denote phase ranges. The vertical bars are error bars.}
\end{figure}

In interval IV which is covered by the slow rise in count rate in NRL11, both the LFQPO frequency and power-law index are relatively steady while the radius of the inner disk increases significantly (Figure \ref{fig:lan.LFQPO}; Figure 7 in NRL11). NRL11 argued that the local Eddington limit in the inner disk is responsible for the slow rise during which the radius of the inner disk increases with luminosity at the constant temperature of the inner disk. The fact that the LFQPO frequency and the power-law index remain relatively stable coincidentally indicates a possible correlation between the LFQPO and the corona. If the LFQPO originates from the corona, then since the LFQPO absolute amplitude in interval IV increases slightly (Figure \ref{fig:lan.absrms}), we can interpret the slight decline in the LFQPO amplitude in interval IV (Figure \ref{fig:lan.LFQPO}(d)) as the result of the increase of the disk flux (Figure 9 in NRL11).

\begin{figure}
\centerline{\includegraphics[height=10cm,angle=-90]{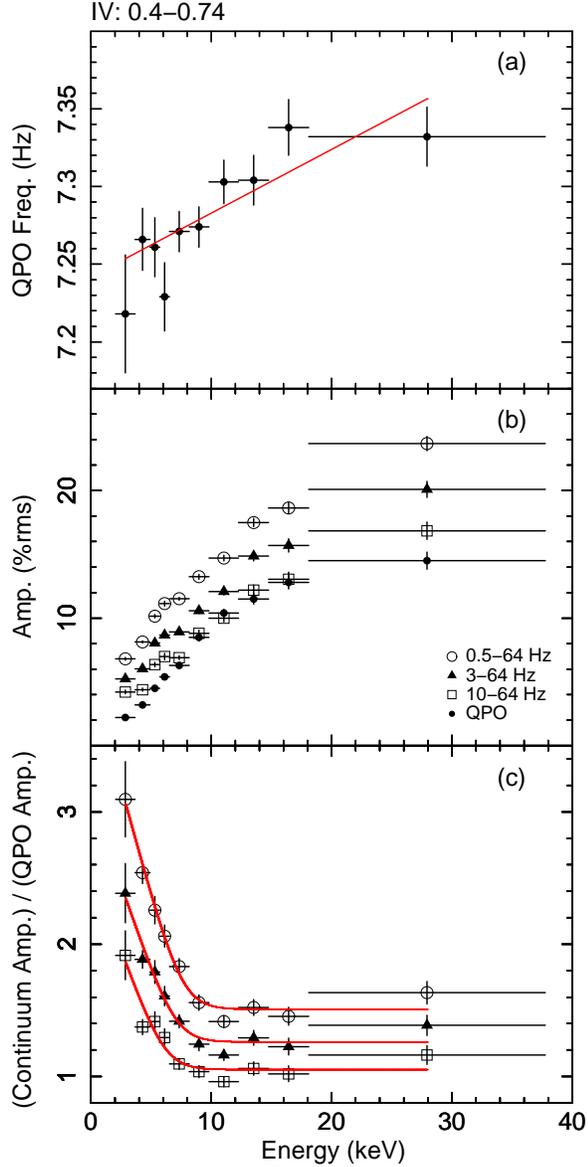}}
\caption{\label{fig:lanRms2e2Fre} (a) the LFQPO frequency spectrum in interval IV. The LFQPO frequency is positively correlated with the photon energy. The points are fitted with Least Squares. (b) the 0.5--64 Hz, 3--64 Hz, 10--64 Hz continuum amplitude spectra and the LFQPO amplitude spectrum in interval IV. As photon energy increases, these amplitudes increase and then flatten out. (c) The ratios of the 0.5--64 Hz, 3--64 Hz, 10--64 Hz continuum amplitudes to the LFQPO amplitude. As photon energy increases, the ratios decrease rapidly and then smoothly level off. The points can be well fitted with function (1) in \citet{Shaposhnikov07}. The open circles, triangles, squares represent the 0.5--64 Hz, 3--64 Hz, 10--64 Hz continua, respectively. The filled circles represent the LFQPO. The horizontal bars denote energy bands. The vertical bars are error bars.}
\end{figure}

In order to further test whether the LFQPO is from the corona, we investigate the spectra of the variability parameters in interval IV. It is found that the LFQPO frequency is positively correlated with photon energy (Figure \ref{fig:lanRms2e2Fre}(a)), which is similar to the behavior of some LFQPOs from GRS~1915+105 during more steady states \citep{Qu10, Yan12}. It was assumed that the inner regions of the accretion flow have a harder spectrum while the outer regions have a softer spectrum \citep[e.g.,][]{Revnivtsev99,Kotov01,Ingram11,Ingram12}. If this assumption is correct, then any satisfying model for the LFQPO should predict that the higher centroid frequency part of the LFQPO is related to the inner part of the flow, while the lower centroid frequency part of the LFQPO is related to the outer part of flow. The Lense--Thirring precession of the hot coronal flow within a truncated disc \citep[and reference therein]{Done07} can produce the LFQPO, with a frequency that is inversely correlated with the character radius of the hot flow \citep{Ingram09}. Then, the Lense--Thirring precession model of the LFQPO seems to have the potential to explain the observed spectrum of the LFQPO frequency. However, although the model could possibly explain the positive correlation between the LFQPO frequency and the photon energy, it is unclear whether it alone can explain the observed negative and intermediate ones \citep{Qu10, Yan12}.

It is also found that as photon energy increases, the LFQPO amplitude increases and then flattens (Figure \ref{fig:lanRms2e2Fre}(b)). If the LFQPO is related to the corona, then the upward trend in the LFQPO amplitude might be a representation of the fractional spectrum of the corona versus the total. The flattening of the amplitude spectrum would then reflect the energy where the disk contribution is negligible. 

It seems impossible that the LFQPO originates from both the disk and the corona simultaneously, due to their essential differences in property and location. Besides, if the LFQPO is produced by the interaction of the disk with the corona, then it is hard to explain why the frequency of the LFQPO produced by the interaction would increase when the radius of the inner disk increases and the frequency of Keplerian motion in the disk decreases. It is therefore very likely that the LFQPO is produced in the corona.

The origin of aperiodic X-ray variability from accretion flows is not very clear. It might be produced by variations propagating in the accretion flow \citep[e.g.,][]{Lyubarskii97, Kotov01, Arevalo06, Titarchuk07, Gierlinski08, Wilkinson09, Uttley11}. Due to the dissipation or filter effect of the flow, all of the higher frequency (e.g., $>$ several Hz) variations created in the outer part of the flow (the disk) may be smoothed out, and the observed higher frequency variations might predominantly comes from the inner coronal flow \citep[see, e.g.,][]{Revnivtsev99, Nowak99, Done07, Titarchuk07, Gierlinski08, Wilkinson09, Ingram11, Heil11}. In this work, we find that the ratios of the continuum amplitudes to the LFQPO amplitude are roughly constants at photon energy $\gtrsim10$keV (Figure \ref{fig:lanRms2e2Fre}(c)), indicating that the continuum in the higher energy band and the LFQPO might originate from the same region of the accretion flow, e.g., the corona. \citet{Ingram09} showed that a hot flow can produce a continuum spectrum and a LFQPO simultaneously. As photon energy decreases, the continuum/LFQPO amplitude ratios increase rapidly (Figure \ref{fig:lanRms2e2Fre}(c)). Assuming that both the continuum in a certain frequency band and the LFQPO are produced in the corona and they have the same amplitude ratio in both the lower and higher energy bands, this hints that the excess continuum power in the lower energy band might be from another part of the flow, e.g., the disk. Then it seems that the disk has high-frequency ($> 10$ Hz) variations (see Figure \ref{fig:lanRms2e2Fre}(c)), which could be explained by X-ray heating of the disk by the varying corona emission \citep[see, e.g.,][]{Wilkinson09}.

There is no obvious LFQPO in interval I, during which there is a hard pulse identified by NRL11. NRL11 argued that some material collides with the hot corona after it has been ejected from the inner disk due to disk instability during the hard pulse phase. Based on the presumption that the LFQPO is produced in the corona, the absence of the LFQPO may be caused by the violent disturbance in the corona owing to the collision.

The LFQPO frequency decreases smoothly during intervals II and III, which cover NRL11's hard X-ray tail phase during which a short-lived jet is said to be produced. Then, the jet seems to be independent of the LFQPO considering that it has no influence on the LFQPO frequency based on the smooth evolution of the LFQPO frequency (Figure \ref{fig:lan.LFQPO}(c)). However, it seems to have reduced the LFQPO amplitude considering the dips in the profiles of the LFQPO amplitude (Figure \ref{fig:lan.LFQPO}(d)) and of the LFQPO absolute amplitude (Figure \ref{fig:lan.absrms}). The decrease of the LFQPO amplitude may be caused by the increase in flux of a component which is independent of the LFQPO. Besides, it is also possible that the LFQPO amplitude itself has been decreased by a certain process, e.g., more accretion material forms the jet but not the corona.

The LFQPO frequency drops at $\phi=0.92$ (Figure \ref{fig:lan.LFQPO}(c)). NRL11 argued that the disk becomes unstable and the inner disk radius decreases rapidly after phase 0.9 due to radiation pressure. Apart from the steep rise in the power-law index (Figure 7 in NRL11), the drop in the LFQPO frequency might be another signal indicating that the corona has been changed significantly. However, it is unclear why the LFQPO frequency falls to just half of the hypothetical value. The decreases of the LFQPO amplitude and the LFQPO absolute amplitude in interval V might be caused by the softening of the source during the corresponding interval (Figure 2e in NRL11). The increase of the LFQPO amplitude and the steep rise of LFQPO absolute amplitude in interval VI might show the impact of disk instability and the quick decrease of the inner disk radius on the corona.

\begin{figure}
\centerline{\includegraphics[height=12cm,angle=-90]{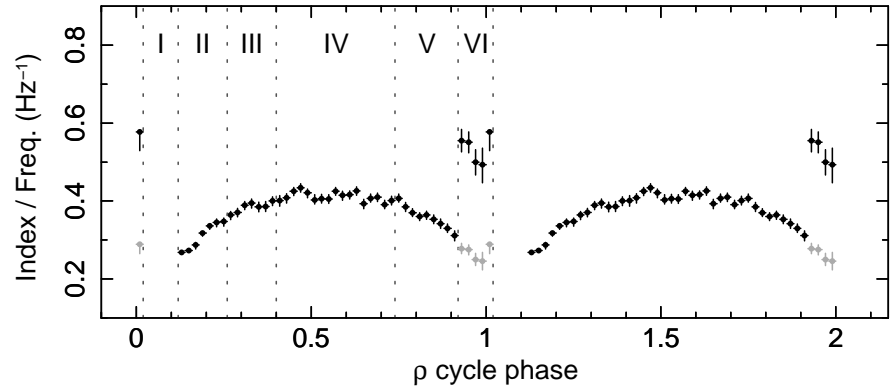}}
\caption{\label{fig:lanGa2Fre} The ratio of the power-law index to the LFQPO frequency as a function of $\rho$ cycle phase. The index is got from Figure 7 in \citet{Neilsen11}. In interval IV, the frequency and the index are both relatively stable, and the ratio is $\sim0.4$. As the frequency increases, the index increases, while the ratio decreases faster. The horizontal bars denote phase ranges. The vertical bars are error bars.}
\end{figure}

The correlation between the spectral index and the LFQPO frequency, which provides us with information about the origin of the LFQPO and the evolution of the source, has been well studied \citep[e.g.,][]{Titarchuk98, Vignarca03, Titarchuk04, Shaposhnikov06, Shaposhnikov07}. It is found that there is a saturation of the spectral index for high values of the LFQPO in some black hole binaries. A transition layer model was established to explain the LFQPO origin and the saturation of the spectral index \citep[e.g.,][]{Titarchuk98, Titarchuk04}. For GRS~1915+105 during the $\chi$ state, the index saturates as the LFQPO frequency is larger than $\sim3$ Hz (Figure 1 in \citet{Shaposhnikov07}). Nevertheless, in the $\rho$ state, the LFQPO frequency is larger than 6 Hz, while the index is positively correlated with the frequency, although there is a trend of saturation (Figure \ref{fig:lanGa2Fre}). \citet{Titarchuk04} explained a similar phenomenon and argued that it occurs when a relatively cold outflow from winds downscatters corona photons and softens the spectrum. Considering that it is hard to tell whether the change in spectral index is caused by the corona change or the wind change, we should be cautious about adopting this argument to interpret our result, though there are some agreements between the evolution of the flux of the Fe XXVI absorbed line and the evolution of the ratio of the spectral index to the LFQPO frequency (Figure 12 in NRL11; Figure \ref{fig:lanGa2Fre}).

In interval IV, the LFQPO frequency spectrum is similar to those of some LFQPOs from GRS~1915+105 during more steady states \citep{Qu10, Yan12}. The LFQPO amplitude spectrum is also similar to that of a 4 Hz QPO from GRS~1915+105 during the $\chi$ state \citep{Zdziarski05}. Both of the similarities suggest that these LFQPOs from GRS~1915+105 during different states might share the same origin.

In summary, we have obtained detailed $\rho$ cycle evolutions of LFQPO parameters. An elaborate comparison analysis is carried out between the results of timing and spectral analyses of GRS~1915+105 during $\rho$ state. It shows that the LFQPO is likely originated from the corona. The similar spectra of variability parameters indicate that the LFQPOs of GRS~1915+105 in very different states seem to be produced from the same mechanism. The continuum / LFQPO amplitude ratio is used to probe the origin of the aperiodic X-ray variability from the accretion flow.

\section*{Acknowledgements}

We thank the anonymous referee for his/her perspective comments which make us greatly improve the quality of the manuscript. We thank Dr. Ming-Yu Ge for his helpful technique discussion and Ms. Qian Yang for her help with the English. This work is supported by CAS (KJCX2-YW-T09), 973 Program (2009CB824800), NSFC (11143013, 11173024, 11203064, and 11203063), and WLFC (XBBS201121, LHXZ201201, and XBBS201123). The research has made use of data obtained from the High Energy Astrophysics Science Archive Research Center (HEASARC), provided by NASA's Goddard Space Flight Center.

\bibliography{yan}

\end{document}